\documentstyle[preprint,pra,floats,aps,psfig]{revtex}

\begin{document}

\title{Coherent control of enrichment and conversion of\\ molecular spin
        isomers}
\author{P.L.~Chapovsky\thanks{E-mail: chapovsky@iae.nsk.su}}
\address{Institute of Automation and Electrometry,\\
          Russian Academy of Sciences, 630090 Novosibirsk, Russia}
         
\date{\today}

\draft
\maketitle

\begin{abstract}
A theoretical model of nuclear spin conversion in molecules
controlled by an external electromagnetic radiation resonant to 
rotational transition has been developed. It has been shown that 
one can produce an enrichment of spin isomers and influence their 
conversion rates in two ways, through coherences and through level
population change induced by radiation. Influence of conversion 
is ranged from significant speed up to almost complete inhibition
of the process by proper choice of frequency and intensity of the 
external field.
\end{abstract}

\vspace{2cm}
\pacs{03.65.-w; 32.80.Bx; 33.50.-j;}

\section{Introduction}

It is well known that many symmetrical molecules exist in Nature only
in the form of nuclear spin isomers \cite{Landau81}.
Spin isomers are important for fundamental science 
and have various applications. They can serve as spin labels,   
influence chemical reactions \cite{Quack77MP,Uy97PRL}, 
or tremendously enhance NMR signals \cite{Bowers86PRL,Natterer97PNMRS}. 
Progress in the spin isomers study depends heavily on available 
methods for isomer enrichment. Although this field has 
been significantly advanced recently (see the review in
\cite{Chap99ARPC}), one needs more efficient enrichment methods. 

Recently two enrichment methods based on optically induced change
of molecular level populations have been proposed 
\cite{Ilichov98CPL,Shalagin99JETPL}. The purpose 
of the present paper is to investigate the isomer enrichment caused
by optically induced coherences in the molecule. 

\section{Quantum relaxation}

\begin{figure}[hb]
\centerline{\psfig
{figure=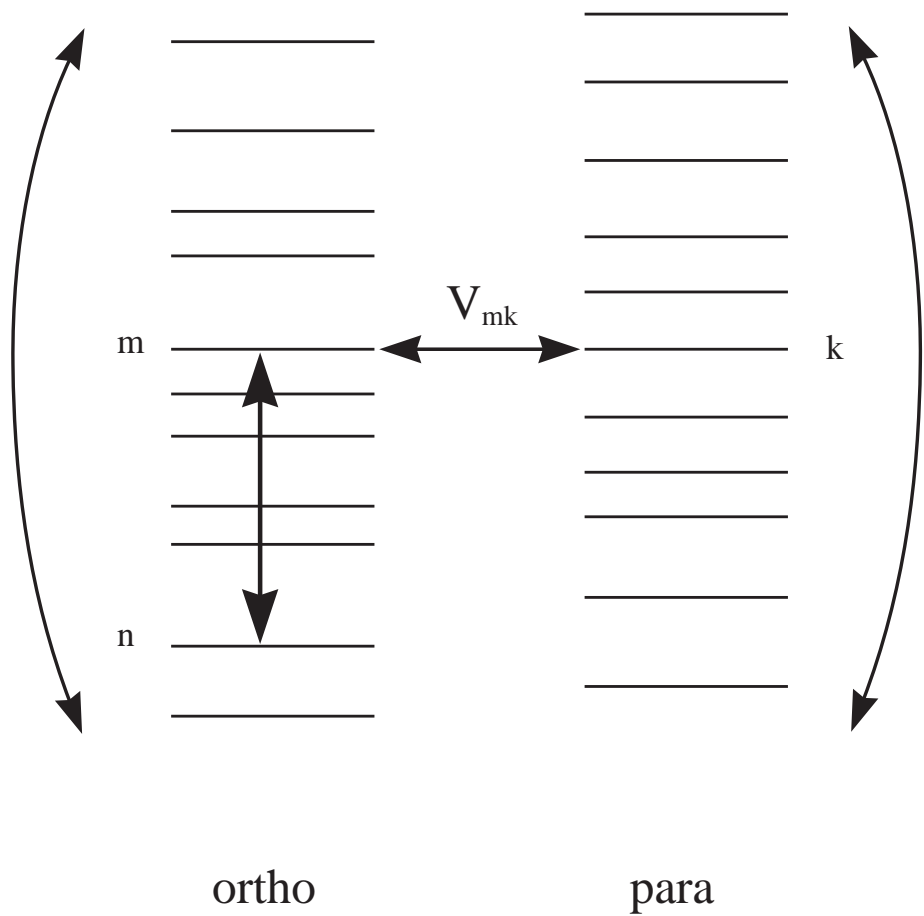,height=12cm}}
\vspace{2cm}
\caption{\sl Molecular ortho and para states. Bent lines indicate
the rotational relaxation. Vertical line shows an optical excitation.
$V_{mk}$ gives the ortho-para state mixing by an intramolecular
perturbation.}
\label{fig1}
\end{figure}

Nuclear spin isomers of molecules were discovered in the late 1920s. The
most known example is the isomers of H$_2$. These isomers 
have different total spin of the two hydrogen nuclei, $I=1$ for
ortho molecules and $I=0$ for para molecules. Symmetrical polyatomic
molecules have nuclear spin isomers too. For example, CH$_3$F can be in
ortho, or in para state depending on the total spin of three hydrogen
nuclei equal 3/2, or 1/2, respectively (see, e.g., \cite{Landau81}).
Different spin isomers are often distinguished also by their rotational quantum
numbers. Consequently, all rotational states of the molecule are separated 
into nondegenerate subspaces of different spin states.
Schematically, these subspaces for the case of two spin states, ortho and
para, are presented in Fig.~\ref{fig1}. (For a moment, an optical 
excitation shown in the ortho space has to be omitted.)

Nuclear spin conversion can be produced by collisions with magnetic particles.
This is a well-known mechanism for the conversion of hydrogen
imbedded in paramagnetic oxygen \cite{Wigner33ZfPC}. 
If a molecule is surrounded by nonmagnetic particles, 
collisions alone cannot change the molecular spin state.
In this case the spin conversion is governed by quantum relaxation  
which can be qualitatively described as follows. 
Let us split the molecular Hamiltonian into two parts,
\begin{equation}
     \hat H = \hat H_0 + \hbar\hat V,
\label{H1}
\end{equation}
where $\hat H_0$ is the main term which has
the ortho and para states as the eigen states (states in Fig.~\ref{fig1});
$\hat V$ is a small {\it intramolecular} perturbation able 
to mix the ortho and para states. 
Suppose that the test molecule was placed initially in the 
ortho subspace. Due to collisions 
the molecule starts undergo fast rotational relaxation 
{\it inside} the ortho subspace. This running up and down
along the ortho ladder proceeds until the molecule reaches  the
ortho state $m$ which is mixed with the para state $k$ by the 
intramolecular perturbation $\hat V$. Then, during the free flight
just after this collision, the perturbation $\hat V$ mixes the para state $k$
with the ortho state $m$. Consequently, the next collision can
move the molecule to other para states and thus localize
it inside the para subspace. Such  mechanism of spin conversion was 
proposed in the theoretical paper \cite{Curl67JCP} (see also \cite{Chap91PRA}). 

Relevance of the described mechanism to actual spin conversion
in molecules is not at all obvious. The problem is that the
intramolecular perturbations, $\hat V$, able to mix ortho and para states are
very weak. They have the order of $10-100$~kHz (hyperfine interactions)
which should be compared with other much stronger
interactions in molecules, or with gas collisions. Nevertheless,
the experimental and theoretical proves have been obtained 
that spin conversion in molecules is indeed governed by quantum relaxation 
\cite{Chap99ARPC,Peters99CPL,Chap00CPL,Chap00PRA}. Although, only three
molecules (CH$_3$F, H$_2$CO and C$_2$H$_4$) have been studied in this
context so far, it is very probable that spin conversion 
in other polyatomic molecules of similar complexity is governed by quantum 
relaxation too. It is useful for the following 
to give a few examples of spin conversion rates. 
Most studied is the spin conversion in $^{13}$CH$_3$F which has the rate,
\begin{equation}
              \gamma_{13}/P = (12.2 \pm 0.6)\cdot 
              10^{-3}~{\text{s}}^{-1}/{\text{Torr}}, 
\label{g13}
\end{equation}
in case of pure CH$_3$F gas \cite{Chap99ARPC}. Spin conversion in
another isotope modification, $^{12}$CH$_3$F, is by almost
two orders of magnitude slower \cite{Chap99ARPC}. Similar slow conversion
was observed in ethylene, $^{13}$CCH$_{4}$, \cite{Chap00CPL},
\begin{equation}
              \gamma_{eth}/P = (5.2 \pm 0.8)\cdot 
              10^{-4}~{\text{s}}^{-1}/{\text{Torr}}.
\label{eth13}
\end{equation}
These data show that spin conversion by quantum relaxation 
is on $9-11$ orders of magnitude slower
than the rotational relaxation, $\nu\sim10^6 - 10^7$~s$^{-1}$/Torr.

\section{Optical excitation}

In this section we will analyze the spin conversion by 
quantum relaxation in the presence of a resonant electromagnetic
radiation. The level scheme is shown in Fig.~\ref{fig1}. In order
to reveal the main features of the phenomenon we will
consider the process in a simplest arrangement. First of all,
we assume that molecular states are nondegenerate and that 
only one ortho-para level pair,
$m-k$, is mixed by the intramolecular perturbation, $\hat V$. 
Monochromatic radiation is chosen to be in resonance 
with the rotational transition $m-n$.  
In this arrangement the molecular Hamiltonian reads,
\begin{equation}
      \hat H = \hat H_0 + \hbar\hat G + \hbar\hat V.
\label{H2}
\end{equation}
New term, $\hat G$, describes the molecular interaction
with the radiation,
\begin{equation}
     \hat G = - ({\bf E}_0\hat {\bf d}/\hbar)\cos \omega_Lt,
\label{G}
\end{equation}
where ${\bf E}_0$ and $\omega_L$ are the amplitude and frequency 
of the electromagnetic wave; $\hat {\bf d}$ is the  
operator of the molecular electric dipole moment. We have neglected  
molecular motion in the operator $\hat G$ because 
homogeneous linewidth of pure rotational transition is usually 
larger than the Doppler width.

Kinetic equation for the density matrix, $\hat\rho$, in the representation 
of the eigen states of the operator $\hat H_0$ has standard form,
\begin{equation}
     d\hat\rho/dt = {\hat S} - i [\hat G + \hat V,\hat\rho], 
\label{r1}
\end{equation}
where ${\hat S}$ is the collision integral.

Molecules in states ($m, n, k$) interacting with the perturbations
$\hat G$ and $\hat V$ constitute only small fraction of the total 
concentration of the test molecules. Thus, one can neglect collisions 
between molecules in these states in comparison with collision 
with molecules in other states. The latter molecules remain almost 
at equilibrium. Consequently, the collision integral, ${\hat S}$, depends 
linearly on the density matrix for the disturbed states $m$, $n$, and
$k$ even in one component gas. Further, we will assume model of strong collisions
for the collision integral. The off-diagonal elements of ${\hat S}$ are
\begin{equation}
     S_{jj'} = -\Gamma\rho_{jj'}; \ \ \ j,j'\in m,n,k;\ \ \ j\neq j'.
\label{str_off}
\end{equation}
Here, $j$ and $j'$ indicate rotational states of the molecule.
The decoherence rates, $\Gamma$, were taken equal for all off-diagonal 
elements of $\hat S$.

Collisions cannot alter molecular spin state in our model. 
It implies that the diagonal elements of $\hat S$ have to be 
determined separately for ortho molecules,
\begin{equation}
           S_{jj}  =  -\nu\rho_o(j) + \nu w_o(j)\rho_o;\ \ \  
           \rho_o = \sum_j\rho_o(j);\ \ \  j\in {\text{ortho}},
\label{str_ortho}
\end{equation}
and for para molecules,
\begin{equation}
           S_{jj}  =  -\nu\rho_p(j) + \nu w_p(j)\rho_p;\ \ \ 
           \rho_p = \sum_j\rho_p(j);\ \ \  j\in {\text{para}}.
\label{str_para}   
\end{equation}
Here $\rho_o$, $\rho_p$ and $w_o(j)$, $w_p(j)$ are the total concentrations
and Boltzmann distributions of ortho and para molecules.
The rotational relaxation rate, $\nu$, was taken equal for 
ortho and para molecules, because kinetic properties 
of different spin species are almost identical. 

One can obtain from Eq.~(\ref{r1}) an equation of change of the 
total concentration in each spin space \cite{Chap91PRA}. For example, 
for ortho molecules one has, 
\begin{equation}
     d\rho_o/dt = 2Re(i\rho_{mk}V_{km}).
\label{ro}
\end{equation}
In fact, this result is valid for any model of collision integral as 
long as collisions do not change the total concentration 
of molecules in each subspace,
i.e., $\sum_j S_{jj}=0$ if $j\in $~ortho, or $j\in $~para.

Let us approximate the density matrix by the sum of zero and 
first order terms over perturbation $\hat V$,
\begin{equation}
     \hat\rho = \hat\rho^{(0)} + \hat\rho^{(1)}.
\label{r2}
\end{equation}
In zero order perturbation theory the ortho and para subspaces
are independent, i.e., $\rho^{(0)}_{mk}=0$. Consequently, equation of change 
(\ref{ro}) is reduced to,
\begin{equation}
     d\rho_o/dt = 2Re(i\rho^{(1)}_{mk}V_{km}).
\label{dro}
\end{equation}
Note that the spin conversion appears in the second order 
of $\hat V$.

Kinetic equations for zero and first order terms of the
density matrix are given as,
\begin{equation}
       d\hat\rho^{(0)}/dt = {\hat S}^{(0)} - i [\hat G,\hat\rho^{(0)}].
\label{ro0}
\end{equation}
\begin{equation}
     d{\hat\rho}^{(1)}/dt  = {\hat S}^{(1)}
        - i [\hat G,\hat\rho^{(1)}] - i [\hat V,\hat\rho^{(0)}].
\label{ro1}
\end{equation}

We start with zero order perturbation theory. In this approximation,
the para subspace remain at equilibrium,
\begin{equation}
     \rho^{(0)}_p(j) = \rho^{(0)}_p w_p(j);\ \ \ j\in {\text{para}}.
\label{rp}
\end{equation} 
Here $\rho^{(0)}_p$ is the total concentration of para molecules.

Density matrix for ortho molecules can be determined from the two equations
which follows from Eqs.~(\ref{str_off}),(\ref{str_ortho}),(\ref{ro0}):
\begin{eqnarray}
 d\rho^{(0)}_o(j)/dt & = & -\nu \rho^{(0)}_o(j) +\nu w_o(j)\rho^{(0)}_o
          + 2Re(i\rho^{(0)}_o(m|n)G_{nm})[\delta_{jm}-\delta_{jn}];  \nonumber \\
 d\rho^{(0)}_o(m|n)/dt & = & -\Gamma \rho^{(0)}_o(m|n) -
                           i G_{mn}[\rho^{(0)}_o(n) - \rho^{(0)}_o(m)].       
\label{sys1}
\end{eqnarray}
We will assume further the rotational wave approximation,
\begin{equation}
     G_{mn} = - Ge^{-i\Omega t};\ \ 
     G\equiv E_0 \overline{d}_{mn}/2\hbar;\ \ 
     \Omega = \omega_L - \omega_{mn}, 
\label{rwa}
\end{equation}
where the line over symbol indicate a time-independent factor. 
Rabi frequency, $G$, is assumed to be real.
Rotational relaxation, $\nu$, and decoherence, $\Gamma$,
are on many orders of magnitude faster than the spin conversion.
It allows to assume stationary regime for ortho molecules, thus having, 
$d\rho^{(0)}_o(j)/dt=0$. The substitution,
$\rho^{(0)}_o(m|n)=\overline{\rho}^{(0)}_o(m|n)\exp{(-i\Omega t)}$, transforms
Eqs.~(\ref{sys1}) to algebraic equations which can easily be solved.
Thus one has,
\begin{eqnarray}
\rho^{(0)}_o(j) & = & \rho^{(0)}_o\left[ w_o(j) + \frac{2\Gamma}{\nu}
        \frac{G^2\Delta w}{\Gamma^2_B + \Omega^2}(\delta_{jm}-\delta_{jn})
        \right]; \nonumber \\
\overline{\rho}^{(0)}_o(m|n) & = & \rho^{(0)}_o i 
        G\Delta w\frac{\Gamma + i\Omega}{\Gamma^2_B+\Omega^2};\nonumber \\ 
        \Gamma^2_B & = & \Gamma^2+4\Gamma G^2/\nu;\ \ \ 
        \Delta w \equiv w_o(n)-w_o(m).     
\label{solution0}
\end{eqnarray}

We turn now to the calculation of the first order term, $\rho^{(1)}_{mk}$, 
which has to be substituted into the equation of change (\ref{dro}).
The density matrix element, $\rho^{(1)}_{mk}$, can be found from
the two equations which are derived from Eqs.~(\ref{str_off}),(\ref{ro1}),
\begin{eqnarray}
d\rho^{(1)}_{mk}/dt + \Gamma\rho^{(1)}_{mk} + iG_{mn}\rho^{(1)}_{nk} & = & 
            -iV_{mk} [\rho^{(0)}_o(k)-\rho^{(0)}_o(m)]; \nonumber \\
d\rho^{(1)}_{nk}/dt  + \Gamma\rho^{(1)}_{nk} + iG_{nm}\rho^{(1)}_{mk}  & = &  
             iV_{mk} \rho^{(0)}_o(n|m).    
\label{sys2}
\end{eqnarray}

Substitution,
\begin{equation}
     V_{mk}=\overline{V}e^{i\omega t},\ \ (\omega\equiv\omega_{mk});\ \ 
     \rho^{(1)}_{mk} = \overline{\rho}^{(1)}_{mk}e^{i\omega t};\ \ 
     \rho^{(1)}_{nk} = \overline{\rho}^{(1)}_{nk}e^{i(\omega-\omega_{kn}) t},
\label{subs}
\end{equation}
transforms Eq.~(\ref{sys2}) to algebraic equations from which one 
finds $\rho^{(1)}_{mk}$. Then, Eq.~(\ref{dro}) gives the following equation 
for the total concentration of ortho molecules,
\begin{eqnarray}
     \frac{d\rho_o}{dt}  =  2|\overline{V}|^2 & Re &
     \frac{[\Gamma+i(\Omega+\omega)][\rho^{(0)}_p(k)-\rho^{(0)}_o(m)]-
     iG\overline{\rho}^{(0)}_o(n|m)}{F(\Omega)}; \nonumber \\
     F(\Omega) & \equiv & (\Gamma+i\omega)[\Gamma+i(\Omega+\omega)] + 
     G^2.
\label{prob}
\end{eqnarray}

After an appropriate change of notations, the right hand side of 
Eq.~(\ref{prob}), coincides formally with the solution \cite{Rautian79} 
for the work of weak optical field in the presence of strong optical field.
Strong field splits the upper state $m$ on two, which appears as two roots 
of the denominator of Eq.~(\ref{prob}) being the second order
polynomial on $\omega$. It results in two ortho-para level pairs mixed by 
the perturbation $\hat V$ instead of one pair in the absence of an external field.  
In analogy with the optical case, one can distinguish the isomer conversion caused by 
population effects (terms proportional to the level populations $\rho^{(0)}_p(k)$ 
and $\rho^{(0)}_o(m)$) and by coherences (term proportional to the off-diagonal
density matrix element, $\overline{\rho}^{(0)}_o(n|m)$).

Eq.~(\ref{prob}) describes time dependence of the concentration of
ortho molecules, $\rho_o$, in the second order of $\hat V$.
One can neglect at this approximation small difference between $\rho_o$ and 
$\rho^{(0)}_o$. The density of para molecules can be expressed through the 
density of ortho molecules as, $\rho^{(0)}_p=n-\rho^{(0)}_o$,
where $n$ is the total concentration of the test molecules. Using these
points and zero order solution given by Eqs.~(\ref{rp}),(\ref{solution0}), 
one can obtain final equation of change for ortho molecules,
\begin{equation}
     d\rho_o/dt = n\gamma_{op} - \rho_o\gamma;\ \ \ 
     \gamma\equiv\gamma_{op} + \gamma_{po} + \gamma_n + \gamma_{coh},
\label{final}
\end{equation}
where partial conversion rates were introduced,
\begin{eqnarray}
\gamma_{op}  & = & 2|\overline{V}|^2w_p(k) f(\Omega);\ \ \ \ \ \ \  
       f(\Omega) \equiv Re\frac{\Gamma+i(\Omega+\omega)}
                 {F(\Omega)}; \nonumber \\
\gamma_{po}  & = &  2|\overline{V}|^2w_o(m)f(\Omega);  \nonumber \\
\gamma_n  & = & 2|\overline{V}|^2\frac{G^2\Delta w}{\Gamma^2_B+\Omega^2}
         \frac{2\Gamma}{\nu}f(\Omega) ;\nonumber \\
\gamma_{coh} & = & 2|\overline{V}|^2\frac{G^2\Delta w}
        {\Gamma^2_B+\Omega^2}Re\frac{\Gamma-i\Omega}{F(\Omega)}.      
\label{gammas}
\end{eqnarray}
Here the rates $\gamma_{op}$, $\gamma_{po}$, and $\gamma_n$ are
due to molecular level populations and the rate $\gamma_{coh}$ is
due to coherences. The terms $\gamma_{op}$ and $\gamma_{po}$ 
are the only ones which remain in the absence of an external field.
In the radiation free case ($G=0$) the result (\ref{gammas}) becomes 
identical with the solution given in \cite{Chap91PRA}. 

\section{Enrichment}

Solution to Eq.~(\ref{final}) can be presented as,
$\rho_o = \overline\rho_o + \delta\rho_o\exp(-\gamma t)$,
where time-independent part is given by
\begin{equation}
     \overline\rho_o = n\frac{\gamma_{op}}{\gamma}.
\label{overro}
\end{equation}
Without an external radiation (at the instant $t=0$), the equilibrium
concentration of para molecules is equal to, 
\begin{equation}
     \rho_p(0)=n-\rho_o(0)=n\frac{w_o(m)}{w_p(k)+w_o(m)} = \frac{n}{2}.
\label{rp1}
\end{equation}
For simplicity, the Boltzmann factors in Eq.~(\ref{rp1}) were
assumed to be equal, $w_p(k)=w_o(m)\equiv w$, which implies,
$\gamma_{op}=\gamma_{po}$. External
field produces a stationary enrichment of para molecules.
One can derive from Eqs.~(\ref{overro}),(\ref{rp1}),
\begin{equation}
     \beta \equiv \frac{\overline\rho_p}{\rho_p(0)}-1 = 
     1-2\frac{\gamma_{op}}{\gamma}.
\label{bp}
\end{equation}
An enrichment coefficient, $\beta$, is defined here in such a way that
$\beta=0$ if there is no external electromagnetic field. 
Enrichment of ortho molecules is equal to $-\beta$.
Note, that the enrichment, $\beta$, does not depend on the  magnitude of
intramolecular perturbation $\hat V$. It is the consequence of the
assumption that only one ortho-para level pair is mixed. Enrichment,
$\beta$, depends on the ratio of Boltzmann factors, $w_o(n)/w_o(m)$,
but does not depend on the magnitude of $w_o(n)$ itself. 
In further numerical examples relative difference of the Boltzmann 
factors will be chosen as $w_o(n)=1.2w$. 

One needs to specify a few other parameters in order to investigate
properties of the optically induced enrichment. We will use, where
it is possible, parameters relevant to the spin conversion in $^{13}$CH$_3$F.
Thus the decoherence rate, $\Gamma$, will be chosen equal 6~MHz,
which corresponds to the gas pressure of pure CH$_3$F equal 0.2~Torr
\cite{Chap99ARPC}.  Rotational relaxation, $\nu$, will be chosen 
by one order of magnitude slower than the decoherence rate, $\nu=0.1\Gamma$. 

Expressions for the enrichment, $\beta$, are given in the Appendix.
If ortho-para mixing is performed for a degenerate pair of states $m-k$ 
($\omega=0$), the enrichment of para states, $\beta$, has one peak 
at $\Omega=0$. This peak is determined mainly by population effects.

\begin{figure}[hb]
\centerline{\psfig
{figure=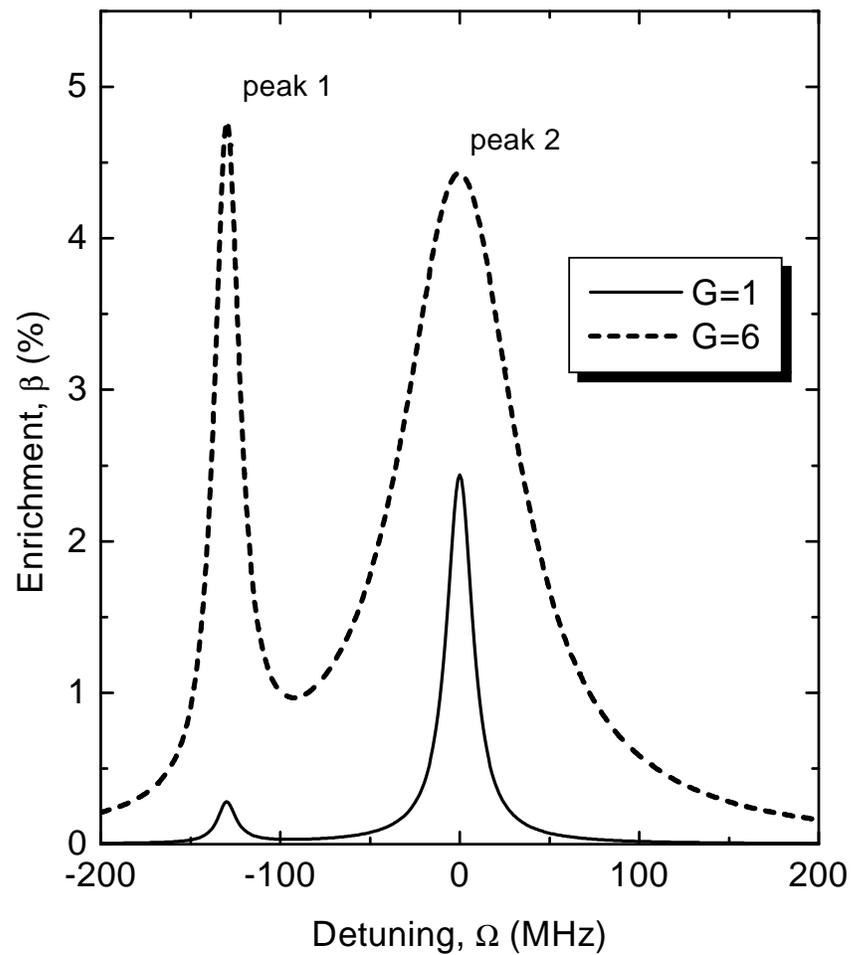,height=20cm}}
\caption{\sl Enrichment of para molecules, $\beta$,
as a function of radiation frequency detuning, $\Omega$, at 
$\omega=130$~MHz and two Rabi frequencies, $G=1$~MHz and $G=6$~MHz.}
\label{fig2}
\end{figure}

\begin{figure}[hb]
\centerline{\psfig
{figure=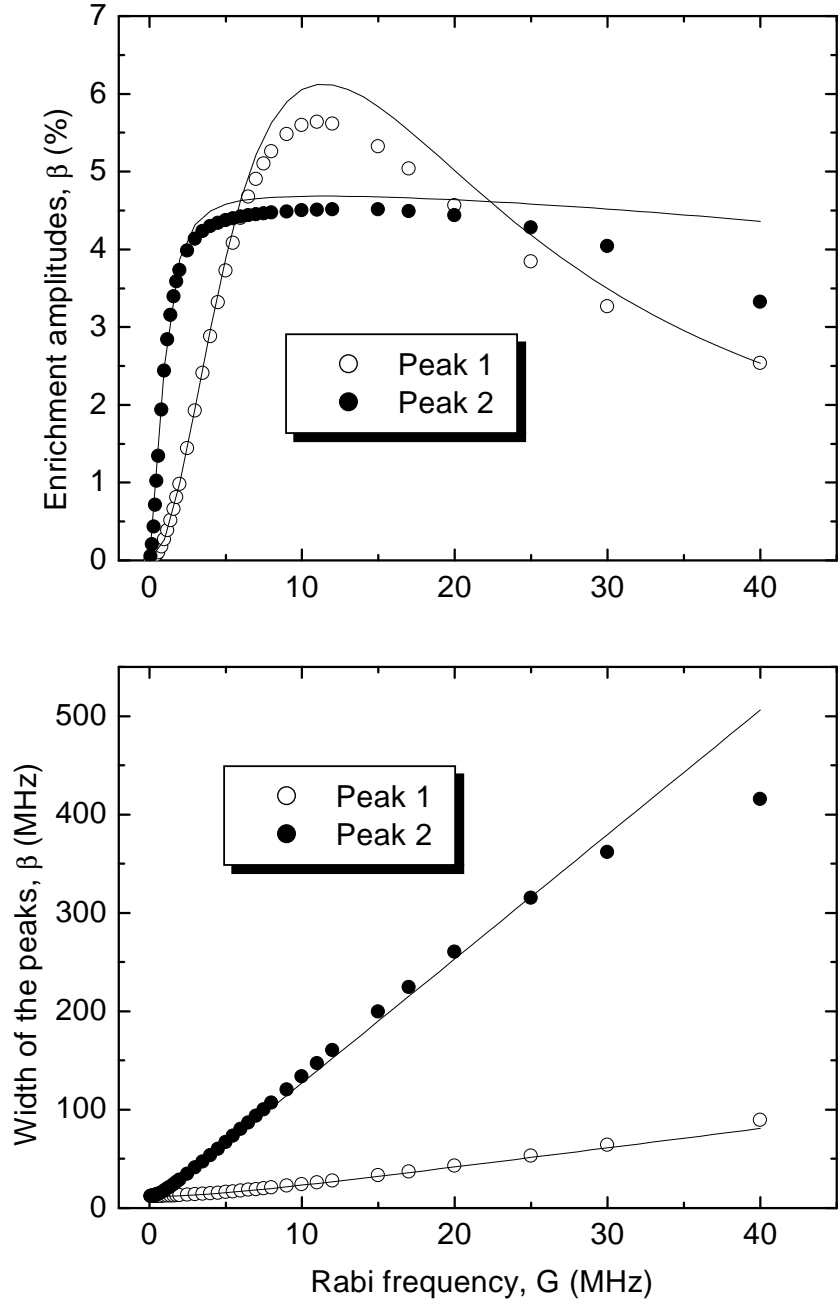,height=20cm}}
\caption{\sl Amplitudes of the peaks in enrichment, $\beta$ (upper panel)
and width (FWHM) of these peaks (low panel) for the case of $\omega=130$~MHz.}
\label{fig3}
\end{figure}

More interesting is the case of nondegenerate states ($\omega\neq0$).
In this case one has two peaks, at $\Omega=-\omega$ and at $\Omega=0$
(Fig.~\ref{fig2}). 
Peak~1 ($\Omega=-\omega$) is due to the coherent effects determined by 
$\gamma_{coh}$. Peak~2 ($\Omega=0$) is mainly due to the optically 
induced level population changes determined by $\gamma_n$.
In the case of well-separated peaks ($\Gamma_B\ll \omega$), the
amplitudes of the peak~1 and peak~2 read,
\begin{eqnarray}
 A_1 & = & \frac{\Delta w}{w}\frac{G^2}{\Gamma^2+G^2}\left[ 
         \frac{1}{4} + \frac{D}{\Gamma^2_B+\omega^2}\right];\ \ \ 
         D\equiv \Gamma^2+5G^2/4-\Gamma G^2/\nu+\omega^2/4; \nonumber \\
 A_2 & = & \frac{\Delta w}{w}\frac{G^2}{\Gamma^2_B}\left[
         \frac{\Gamma}{\nu}-\frac{3}{4} + \frac{D}
         {\Gamma^2_B+G^2+\omega^2}\right].      
\label{A12}
\end{eqnarray}
Amplitude of the peak~2 grows rapidly with $G$ up to
$\beta\simeq 4.5\%$. Amplitude of the peak~1 grows
with $G$ to even bigger value $\beta\simeq 5.5\%$. These data are
shown in Fig.~\ref{fig3} (upper panel) where points are obtained by 
fitting an exact expression (\ref{bp}) by two Lorentzians and solid
curves are given by Eqs.~(\ref{A12}). $\omega$ was chosen equal
130~MHz which corresponds to the ortho-para level gap
in $^{13}$CH$_3$F \cite{Chap99ARPC}. Note, that there is no
optically induced enrichment if $\Delta w=0$.

Widths of the enrichment peaks are given by the expressions,
\begin{equation}
     W_1 = 2\sqrt{\Gamma^2+G^2};\ \ \ W_2 = 2\Gamma_B.
\label{W12}
\end{equation}
The two enrichment peaks experience completely different
power broadening which are shown in Fig.~\ref{fig3} (low panel).
Peak~1 has the width much smaller than the 
Peak~2. Solid curves in Fig.~\ref{fig3} (low panel)
are given by Eqs.~(\ref{W12}). Points are obtained by fitting
the exact expression for enrichment, $\beta$, by two
Lorentzians.

\section{Conversion}

Conversion rate in the presence of an external electromagnetic field
has complicated dependence on radiation frequency detuning, $\Omega$,
and Rabi frequency, $G$. It is convenient to characterize the
conversion rate in relative units,
\begin{equation}
     \gamma_{rel} = \frac{\gamma}{\gamma_{free}} - 1,
\label{grel}
\end{equation}
where $\gamma_{free}$ is the field free conversion rate. 
Similar to the enrichment, $\beta$, this parameter does not
depend on magnitude of the perturbation $\hat V$ and on absolute
values of the Boltzmann factors. 

\begin{figure}[hb]
\centerline{\psfig
{figure=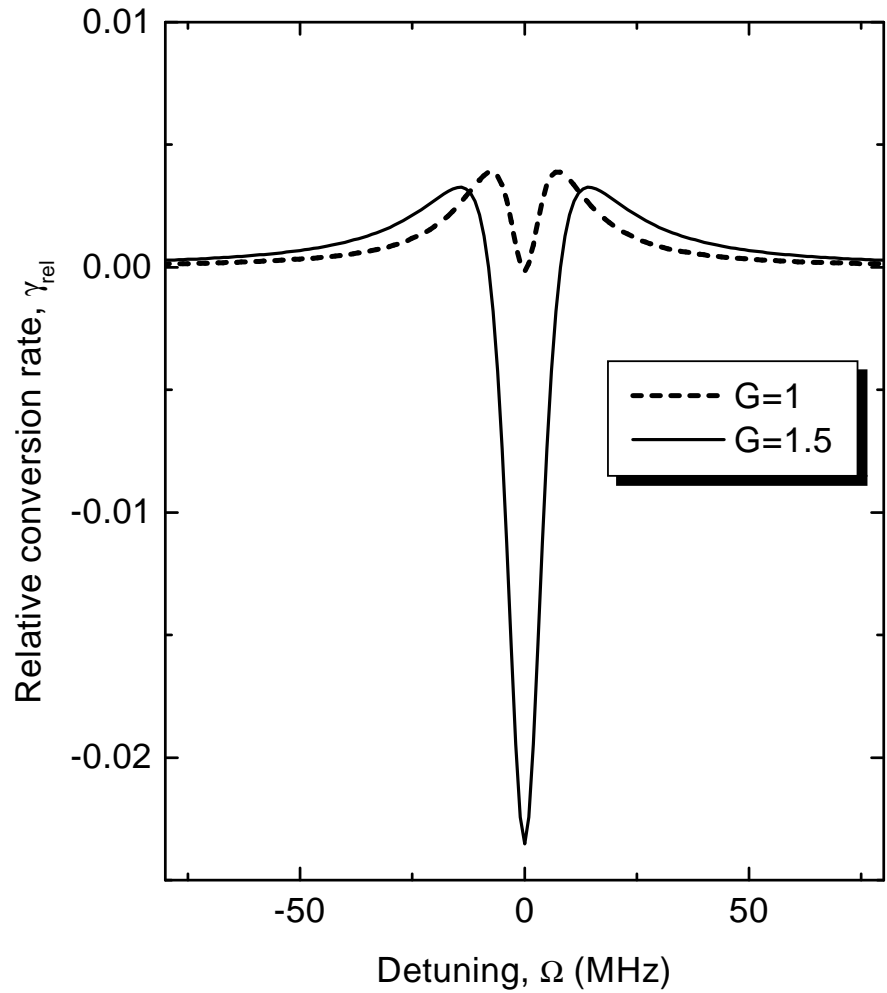,height=20cm}}
\caption{\sl Relative conversion rate, $\gamma_{rel}$, for the
case of degenerate ortho-para states, $\omega=0$ and the two
values of Rabi frequency, $G=1$~MHz and $G=1.5$~MHz.}
\label{fig4}
\end{figure}

Expressions for the conversion rate, $\gamma_{rel}$, are given
in the Appendix.
In the case of degenerate ortho-para level pair $m-k$ ($\omega=0$),
$\gamma_{rel}$ has narrow negative structure at $\Omega=0$ (Fig.~\ref{fig4}).
Amplitude of this dip grows rapidly with increasing $G$. If
$G\gg\Gamma\gg\nu$, and all Boltzmann factors have the 
same order of magnitude, the conversion rate at $\Omega=0$ is given by
\begin{equation}
     \gamma_{rel} \sim \frac{\Gamma^2}{G^2}-1.
\label{grel1}
\end{equation}
If $G$ is large, the relative conversion rate, $\gamma_{rel}\simeq-1$,
which corresponds to $\gamma\simeq0$.
Thus the spin conversion can be inhibited by radiation having 
large $G$ and $\Omega=0$.

\begin{figure}[hb]
\centerline{\psfig
{figure=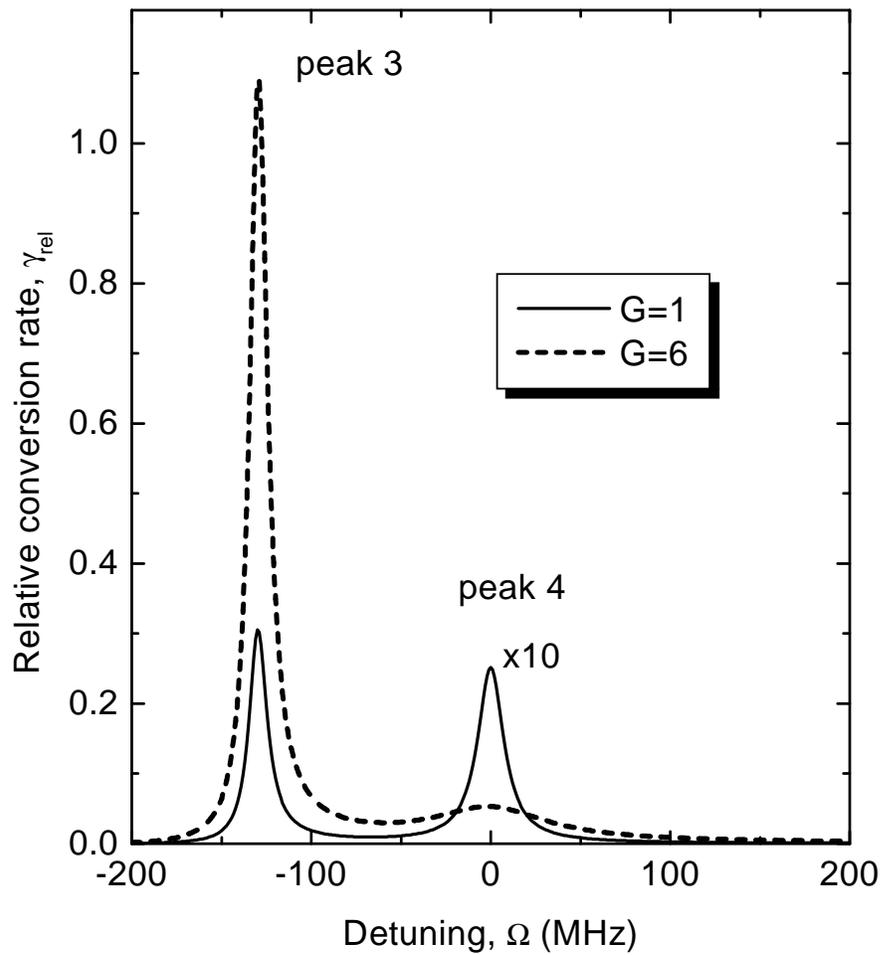,height=20cm}}
\caption{\sl Relative conversion rate, $\gamma_{rel}$, for the
case of nondegenerate ortho-para states ($\omega=130$~MHz) at
two Rabi frequencies, $G=1$~MHz and $G=6$~MHz.}
\label{fig5}
\end{figure}

In the nondegenerate case ($\omega\neq0$) conversion rate, $\gamma_{rel}$,
has two peaks (Fig.~\ref{fig5}). If these peaks are well resolved, 
$\Gamma_B\ll\omega'$, conversion rate is given by,
\begin{equation}
     \gamma_{rel} = \frac{\Gamma'\Gamma^{-1}G^2}{\Gamma'^2+(\Omega+\omega')^2} 
                 +\frac{\Delta w}{w}\frac{G^2}{\Gamma^2_B + \Omega^2}
                 \left(\frac{\Gamma}{\nu}-\frac{1}{2}\right),
\label{grel2}
\end{equation}
where new parameters are determined as,
\begin{equation}
     \Gamma'\equiv\Gamma\left(1+\frac{G^2}{\Gamma^2+\omega^2}\right); \ \ \ 
     \omega'\equiv\omega\left(1-\frac{G^2}{\Gamma^2+\omega^2}\right).
\label{gom}
\end{equation}

\begin{figure}[hb]
\centerline{\psfig
{figure=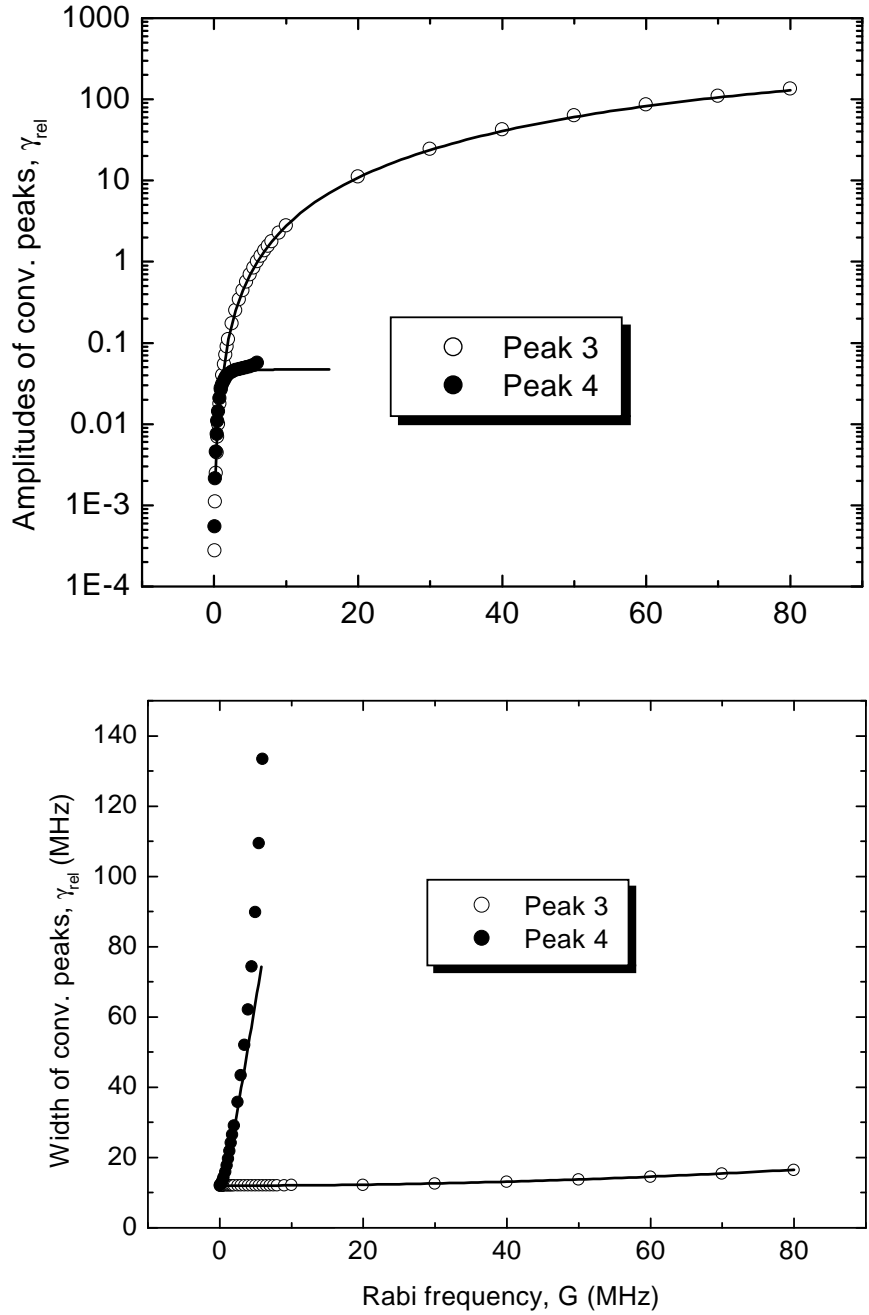,height=20cm}}
\caption{\sl Amplitudes of the conversion peaks, $\gamma_{rel}$, 
(upper panel) and widths (FWHM) of these peaks (low panel) for the
nondegenerate case, $\omega=130$~MHz.}
\label{fig6}
\end{figure}

The two peaks in the conversion have Lorentzian profiles and
amplitudes determined by the expressions,
\begin{equation}
     A_3 = \frac{G^2}{\Gamma'\Gamma}; \ \ \ 
     A_4 = \frac{\Delta w}{w}\frac{G^2}{\Gamma^2_B}
           \left(\frac{\Gamma}{\nu}-\frac{1}{2}\right).
\label{A34}
\end{equation}
These amplitudes are shown in Fig.~\ref{fig6} (upper panel)
by solid curves. Points are obtained by fitting the exact solution
by two Lorentzians. In this examples, $\omega$ was chosen equal 130~MHz. 
Peak~4 at $\Omega=0$ is proportional to the ratio of Boltzmann
factors. This peak does not grow significantly with $G$. 
The peak~3 at $\Omega=-\omega'$ almost does not depend 
on Boltzmann factors and at large $G$ grows up to 
$\gamma_{rel}=(\omega/\Gamma)^2$. Thus strong electromagnetic 
field can speed up the conversion significantly, viz., 
by two orders of magnitude in our numerical example.

Widths of the two peaks in the conversion rate are determined 
by the equations,
\begin{equation}
     W_3 = 2\Gamma';\ \ \  W_4 = 2\Gamma_B,
\label{W34}
\end{equation}
and have very different field dependences.
In fact, the peak~4 at $\Omega=0$ is broadened even faster
than $2\Gamma_B$. Peak~3 at $\Omega=-\omega$ has almost no power 
broadening, if $G\ll\omega$ (Fig.~\ref{fig6}, low panel). Solid curves
in Fig.~\ref{fig6}, low panel, corresponds to the Eqs.~(\ref{W34}).
Points are obtained by fitting the exact solution
by two Lorentzians.

\section{Conclusions}

We have shown that an external resonant radiation can influence
spin isomer conversion in two ways, through level populations and
through optically induced coherences. The coherences 
introduce in the process new and interesting features. In many
cases, the coherences play more important role than the level populations.
This analysis have been performed in a simplest arrangement 
in order to reveal the main features of the phenomenon. 

Optically induced coherences introduce extra resonances both in 
enrichment and in conversion frequency dependences.
These new resonances are important for future 
experimental realizations of the optical control of isomer
conversion. First, they give convenient opportunity to find
coincidences between molecular transitions and available sources
of powerful radiation. Second, an observation of the phenomenon will
be easier also because electromagnetic radiation can significantly
speed up the conversion. Thus steady state enrichment can be
achieved much faster than without field. It
allows to work at low gas pressures where described above
effects can be achieved at smaller radiation intensity.

Another advantage is that one can use resonance at which there is
no large radiation absorption and, consequently, no significant
level population change by radiation. It should decrease some spurious 
effects, like molecular resonance exchange. The latter effect can
cause serious problems in realization of the optically induced
enrichment by population effects \cite{Ilichov98CPL,Shalagin99JETPL}.

\section{Appendix}

Here we give expressions for the enrichment, $\beta$, and conversion rate,
$\gamma_{rel}$. Enrichment of para molecules is given by Eq.~(\ref{bp}).
For equal Boltzmann factors, $w_o(m)=w_p(k)\equiv w$, one has
\begin{equation}
     \beta = 1 - \left(1+\frac{\gamma_n+\gamma_{coh}}{2\gamma_{op}}
                 \right)^{-1},
\label{bp1}
\end{equation}
and an approximate expression in case of small enrichment,
\begin{equation}
     \beta  \simeq \frac{\gamma_n+\gamma_{coh}}{2\gamma_{op}}.
\label{bp2}
\end{equation}
Using Eqs.~(\ref{gammas}), this expression can be reduced to,
\begin{eqnarray}
     \beta  \simeq & \frac{\Delta w}{w} & \left[
                      \frac{1/4}{\Gamma^2+G^2+(\Omega+\omega)^2}+
                      \frac{\Gamma/\nu-3/4}{\Gamma^2_B+\Omega^2}+
                      \frac{D}{(\Gamma^2_B+\Omega^2)
                      [\Gamma^2+G^2+(\Omega+\omega)^2]}\right]; \nonumber \\
                 D &  \equiv & \Gamma^2 + 5G^2/4 - \Gamma G^2/\nu
                               + \omega^2/4.      
\label{bp3}
\end{eqnarray}
There are two peaks in enrichment, at $\Omega=-\omega$ and at $\Omega=0$.
These peaks have Lorentzian shape if $\omega\gg\Gamma_B$. In the 
degenerate case ($\omega=0$) there is one peak of complicated form at 
$\Omega=0$.

The spin conversion rate, $\gamma_{rel}$, is defined as,
\begin{equation}
     \gamma_{rel} = \frac{\gamma}{\gamma_{free}}-1.
\label{grel3}
\end{equation}
Using Eqs.~(\ref{gammas}) one can obtain after straightforward
calculations the following expression,
\begin{eqnarray}
\gamma_{rel}  = && g \frac{(1+g)(\omega^2-\Gamma^2)+
                        2\omega(\Omega+\omega')}
                   {\Gamma'^2+(\Omega+\omega')^2} +  \nonumber \\
               && \frac{\Delta w}{w} \frac{G^2}{\Gamma^2_B+\Omega^2}               
                  \left(\frac{\Gamma}{\nu}-\frac{1}{2}\right)\left[1+
                  g\frac{(1+g)(\omega^2-\Gamma^2)+2\omega(\Omega+\omega')}
                  {\Gamma'^2+(\Omega+\omega')^2}\right]+ \nonumber  \\
               && \frac{\Delta w}{2w}\frac{G^2}{\Gamma^2_B+\Omega^2}
                  \frac{(1+g)(2\Gamma^2+\omega^2)-
                                \omega(\Omega+\omega')}
                  {\Gamma'^2+(\Omega+\omega')^2};  \nonumber  \\
               && g\equiv\frac{G^2}{\Gamma^2+\omega^2};\ \ \ 
                  \Gamma'\equiv\Gamma(1+g); \ \ \ 
                  \omega'\equiv\omega(1-g).     
\label{grel4}
\end{eqnarray}
Thus the conversion rate has two peaks situated at $\Omega=-\omega'$
and at $\Omega=0$. Position of the peak at $\Omega=\omega'$ depends
on the field intensity.

In the case of degenerate ortho-para level pair ($\omega=0$), frequency 
dependence of $\gamma_{rel}$ has complicated shape with a dip in the center at
$\Omega=0$ (Fig.~(\ref{fig4})). The whole structure is described by the
expression, 
\begin{equation}
  \gamma_{rel} = -\frac{(1+g)G^2}{\Gamma'^2+\Omega^2} + 
                \frac{\Delta w}{w}\frac{G^2}{\Gamma^2_B+\Omega^2}
      \left[\left(\frac{\Gamma}{\nu}-\frac{1}{2} \right)
      \left(1-\frac{(1+g)G^2}{\Gamma'^2+\Omega^2}\right) + 
      \frac{\Gamma'\Gamma}{\Gamma'^2+\Omega^2}\right].
\label{grel5}
\end{equation}
   
\section*{Acknowledgments}

This work was made possible by financial support from the  
Russian Foundation for Basic Research (RFBR), grant No. 98-03-33124a

\newpage

\newpage

\newpage

\newpage

\newpage

\newpage

\end{document}